\documentclass[prd,aps,showpacs,twocolumn]{revtex4}
\usepackage{graphicx}
\begin{document}

\title{Extra-galactic magnetic fields and the second knee in the
cosmic-ray spectrum}

\author{Martin Lemoine} \affiliation{Institut d'Astrophysique de
Paris,\\ UMR7095 CNRS, Universit\'e Pierre \& Marie Curie, \\
98 bis boulevard Arago, F-75014 Paris, France}
\date{\today}
\begin{abstract}
Recent work suggests that the cosmic ray spectrum may be dominated by
Galactic sources up to $\sim 10^{17.5}\,$eV, and by an extra-Galactic
component beyond, provided this latter cuts off below the transition
energy. Here it is shown that this cut-off could be interpreted in
this framework as a
signature of extra-galactic magnetic fields with equivalent average
strength $B$ and coherence length $l_{\rm c}$ such that $B\sqrt{l_{\rm
c}}\sim 2-3\cdot10^{-10}\,$G$\cdot$Mpc$^{1/2}$, 
assuming $l_{\rm c}<r_{\rm L}$ (Larmor radius at $\lesssim 10^{17}\,$eV) and
continuously emitting sources with density $10^{-5}\,$Mpc$^{-3}$. The
extra-Galactic flux is suppressed below $\sim 10^{17}\,$eV as the
diffusive propagation time from the source to the detector becomes
larger than the age of the Universe.
\end{abstract}
\maketitle


\section{Introduction} 

Recent developments, both experimental
and theoretical, have significantly broadened the landscape of
ultra-high energy cosmic ray phenomenology. The High Resolution Fly's
Eye experiment has reported the detection of a high energy cut-off
$\sim 10^{20}\,$eV~\cite{HiRes}, as would be expected from a
cosmological population of sources. This experiment has also observed
that the chemical composition is dominated by protons down to $\sim
10^{18}\,$eV, and by heavy nuclei further below, in agreement with
preliminary KASCADE data~\cite{KASCADE}. This and the steepening of
the cosmic-ray spectrum at the ``second knee'' $\sim 10^{17.5}\,$eV
suggest the disappearance of the low-energy (heavy nuclei) component
and the nearly simultaneous emergence of a high-energy (proton)
component.  On theoretical grounds, it has been observed by Berezinsky
{\it et al.}~\cite{BGG1} that a cosmological distribution of sources
producing a single powerlaw could fit the high energy part of the
cosmic-ray spectrum from the second knee up to the cut-off at
$10^{20}\,$eV, including the dip of the ''ankle'' $\sim
10^{18.5}$eV. In light of these results, it is thus tempting to think
that the cosmic-ray spectrum consists of only two main components: one
Galactic, another extra-galactic, with the transition around the
second knee. 

  There are alternative views, admittedly, in which the Galactic
component dominates the all-particle spectrum up the 
ankle~\cite{WW04}; this latter feature would then mark the emergence of 
the extra-galactic component rather than the signature of pair 
production as in Ref.~\cite{BGG1}. This issue will be hopefully 
settled by ongoing and future cosmic ray experiments,
through more accurate composition and anisotropy measurements. 
The discussion that follows assumes that the interpretation of 
Berezinsky and coworkers~\cite{BGG1} is correct, namely, that 
the transition between the Galactic and extra-galactic cosmic ray
component arises at the second knee.

  This model then requires to impose a low-energy cut-off on the
extra-galactic spectrum around $10^{18}\,$eV in order to not
overproduce the flux close to the second knee. The exact position
of this cut-off as well as the spectral slope below it must be tuned
to how the Galactic component extends above the knee~\cite{BGG2}.

  The objective of the present work is to demonstrate that this
cut-off could be interpreted as a signature of extra-galactic magnetic
fields with average strength $B$ and coherence length $l_{\rm c}$ such
that $B\sqrt{l_{\rm c}}\sim 2-3\cdot 10^{-10}\,$G$\cdot$Mpc$^{1/2}$, 
assuming $l_{\rm c}< r_{\rm L}$ Larmor radius at $\lesssim
10^{17}\,$eV and continuously emitting sources with density
$10^{-5}\,$Mpc$^{-3}$. In this picture, the extra-Galactic spectrum
shuts off below $10^{17}\,$eV as the diffusion time from the closest
sources becomes larger than the age of the Universe. The first knee is
viewed here as the maximal injection energy for protons at the
(Galactic) source. 

  The existence of extra-galactic magnetic fields is of importance to
various fields of astrophysics, including ultra-high energy cosmic ray
phenomenology, but very little is known on their origin, on their
spatial configuration and on their amplitude~\cite{W03}. The 
upper limits on $B\sqrt{l_{\rm c}}$ from Faraday observations lie some
two orders of magnitude above the value suggested here. In the present
framework, experiments such as KASCADE-Grande~\cite{KASCADE} could
probe these magnetic fields thanks to accurate measurements of the
spectrum and composition in the range $10^{16}\rightarrow10^{18}\,$eV.
\smallskip

\section{Particle propagation} 

The main effect of
extra-galactic magnetic fields on $\sim 10^{17}\,$eV particles is as
follows.  In a Hubble time, these particles travel by diffusing on
magnetic inhomogeneities a linear distance $d \sim
\left(cH_0^{-1}l_{\rm scatt}\right)^{1/2}\simeq 65\,{\rm Mpc}\,(l_{\rm
scatt}/1\,{\rm Mpc})^{1/2}$, with $l_{\rm scatt}$ the scattering
length of the particle. If $d$ is much smaller than the typical
source distance, the particle cannot reach the detector in a Hubble
time; since $l_{\rm scatt}$, hence $d$, increases with
increasing energy, this produces a low-energy cut-off in the
propagated spectrum. Current data at the highest energies, notably the
clustering seen by various experiments, suggests a cosmic-ray source
density $n\sim 10^{-6}-10^{-5}\,$Mpc$^{-3}$~\cite{ns}, which
corresponds to a source distance scale $\sim 50-100\,$Mpc. Hence
$l_{\rm scatt}\lesssim 0.3-1\,$Mpc at $10^{17}\,$eV would shut off the
spectrum below this energy~\cite{ILS01}.

To be more quantitative one has to calculate the propagated spectrum
and compare it to the observed data. The low energy part ($\lesssim
10^{18}\,$eV in what follows) of the extra-galactic proton spectrum
diffuses in the extra-galactic magnetic field since the scattering
length $l_{\rm scatt} \ll d$ ($d$ typical source distance). In
contrast, particles of higher energies ($\gtrsim 10^{19}\,$eV in what
follows) travel in a quasi-rectilinear fashion, meaning that the total
deflection angle $\theta_{\rm rms}\ll 1$, 
since $l_{\rm scatt} \gg l_{\rm loss}$, 
where $l_{\rm loss}\lesssim 1\,$Gpc at $E\gtrsim 10^{19}\,$eV is the energy loss
length (which gives an upper bound to the linear distance
across which particles can travel). 

In the diffusion approximation, the propagated differential spectrum
reads (see the Appendix):
\begin{equation}
J_{\rm diff}(E) \, = \, {c\over 4\pi}\int {\rm d}t\sum_{i}\, {e^{-
r_i^2/(4\lambda^2)}\over (4\pi\lambda^2)^{3/2}}\,{{\rm d}E_{\rm
g}(t,E)\over {\rm d}E}\, Q\left[E_{\rm g}(t,E)\right].\label{eq:diff}
\end{equation} 
 The sum carries over the discrete source distribution; $r_i$ is the 
comoving distance to source $i$. Note that a factor $(a_0/a_{\rm e})^{-3}$
in Eq.~(\ref{eq:Green}), with $a_0$ and $a_{\rm e}$ the
scale factor at observation and emission respectively, 
has been absorbed in defining a comoving source density; 
the remaining factor $(a_0/a_{\rm e})\approx {\rm d}E_{\rm g}/ {\rm
d}E$, see below. The
function $E_{\rm g}(t,E)$ defines the energy of the particle at time
$t$, assuming it has energy $E$ at time $t_0$.  This function and its
derivative ${\rm d}E_{\rm g}/ {\rm d}E$ can be reconstructed by
integrating the energy losses~\cite{BGG1}. In the Appendix, it is shown
that Eq.~(\ref{eq:diff}) provides a solution to the diffusion equation
in the expanding space-time under the assumption that the energy loss
of the particle is dominated by expansion losses, which is found to 
be a good approximation for particles with observed energies
$E\lesssim 10^{18}\,$eV. In this case, ${\rm d}E_{\rm g}/ {\rm d}E
\approx E_{\rm g}/E \approx a_0/a_{\rm e}$. Although photopair
and photopion production losses on diffuse backgrounds are negligible
with respect to expansion losses for most of the particle history,
they set the maximum linear distance (hence the
maximum time) across which a particle can travel. The time integral 
in Eq.(\ref{eq:diff}) is indeed bounded by the maximal
lookback time $t$ at which $E_{\rm g}(t,E)=E_{\rm max}$ and by the
minimal lookback time $t_0-t \approx l_{\rm scatt}/c$ necessary to
enter the diffusing regime, taken to be the solution of $r(t) =
\lambda(t,E)$, where $r(t)=\int_{t}^{t_0} {\rm d}t'/a(t')$ is the
comoving light cone distance.  The (comoving) path
length $\lambda$ is defined in  Eq.~(\ref{eq:lambda}) as
$\lambda^2\,=\,\int_{t_{\rm e}}^{t_0} 
{\rm d}t\, a^{-1}(t)D\left[a_eE_e/a(\eta)\right]$, with 
$E_e =E_{\rm g}(t_e,E)$ the energy at injection. The physical
meaning of $\lambda$ is that of a typical distance traveled by
diffusion, accounting for energy losses.

The injection spectrum extends from some minimum energy 
($\lesssim10^{16}\,$eV in the present
model) up to $E_{\rm max}=10^{22}\,$eV (the exact value is of little 
importance here).  The function
$Q(E_{\rm g}) = K (E_{\rm g}/E_{\rm max})^{-\gamma}$ gives the
emission rate per source at energy $E_{\rm g}$, $K$ a normalization
factor such that $\int {\rm d}E\, EQ(E)=L$, with $L$ the total
luminosity, which is assumed to scale as the cosmic star formation
rate from~\cite{SH04}. This theoretical star formation rate history 
agrees with existing data at moderate redshifts and provides an argued
prediction for higher redshifts. It also fits
in nicely the constraints of the diffuse supernova neutrino 
background~\cite{SBWZ05} which would be violated by more steeply
evolving star formation rates. The choice of the star formation rate is not 
crucial to the present analysis since the exponential cut-off due
to the magnetic horizon dominates the effect of the star formation
history on the low energy part of the spectrum. 

 Only continuously emitting sources are considered
here, although the effect of a finite activity timescale for the source is discussed
further below. The cosmological evolution of the magnetic field has been
neglected for simplicity; if the diffusion coefficient  
depends explicitly on time $t$ the solution Eq.~(\ref{eq:diff})  remains exact. 

At higher energies, the propagated spectrum is given by:
\begin{equation}
J_{\rm rect}(E)\,=\, {1\over 4\pi}\sum_i\, {1\over 4 \pi r_i^2}\,
{{\rm d}E_{\rm g}(t_i,E)\over {\rm d}E}\,Q\left[E_{\rm
g}(t_i,E)\right],\label{eq:rect}
\end{equation}
and $t_i$ in Eq.~(\ref{eq:rect}) is related to $r_i$ by
$r_i=\int_{t_i}^{t_0}\,{\rm d}t'/a(t')$; $r_i$ should not exceed
$\lambda(t_i,E)$, beyond which point motion must have become
diffusive.

The Galactic cosmic ray component is modeled as follows. Supernovae
are accepted as standard acceleration sites, yet it is notoriously
difficult to explain acceleration up to a maximal energy $\sim
10^{18}\,$eV. Thus it is assumed that the knee sets the maximal
acceleration energy for Galactic cosmic rays: in this conservative
model, the spectrum of species $i$ with charge $Z$ takes the form
$j_Z(E) \approx (E/E_Z)^{-\gamma_i}\exp(-E/E_Z)$, with $\gamma_i \sim
2.4-2.7$ a species dependent spectral index, $E_Z = Z\times E_{\rm p}$
the location of the knee, with $E_{\rm p}\approx
2\cdot10^{15}\,$eV~\cite{KASCADE}. This scenario is consistent with
preliminary KASCADE data.

  Datasets from the most recent experiments are considered here:
KASCADE $10^{15}\rightarrow 10^{17}\,$eV~\cite{KASCADE}, Akeno
$10^{15}\rightarrow 10^{18.6}\,$eV~\cite{Akeno}, AGASA
$10^{18.5}\rightarrow 10^{20.5}\,$eV~\cite{AGASA}, HiRes
$10^{17.3}\rightarrow 10^{20}\,$eV~\cite{HiRes} and Fly's Eye
$10^{17.3}\rightarrow 10^{20}\,$eV~\cite{FE}. Akeno is not recent but
it is the only experiment whose data bridge the gap between the knee
and the ankle.  These experiments use different techniques and their
results span different energy ranges, hence the data do not always
match. A clear example is the discrepancy between HiRes and AGASA at
the highest energies. In the following, the high energy datasets have
been split in two groups: one with HiRes and Fly's Eye, the other with
Akeno and AGASA. The flux of the extra-galactic component was given
two possible normalization values in order to accomodate either of
these datasets, while the flux of the low energy part is scaled to the
recent KASCADE data. A more robust
comparison of this model with the data will be possible with the
upcoming results of the KASCADE-Grande experiment covering the range
$\sim 10^{16}\,\rightarrow 10^{18}\,$eV~\cite{KASCADE}.

\begin{figure}[t]
      \centering
      \includegraphics[width=0.5\textwidth,clip=true]{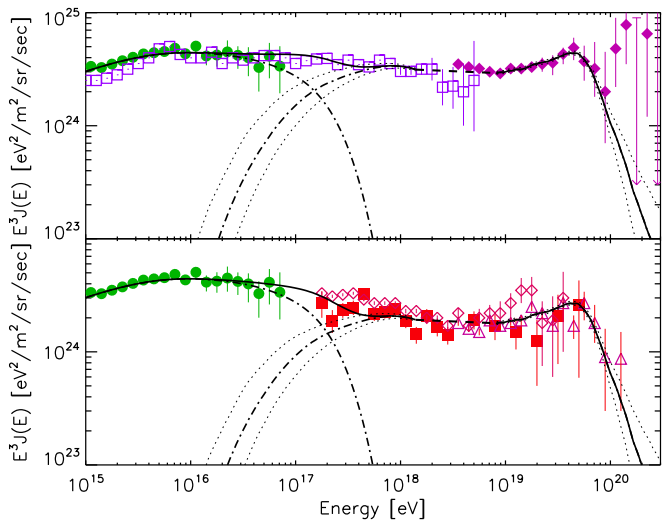}
      \caption[...]{Open squares: Akeno; filled circles: KASCADE;
      filled diamonds: AGASA; filled squares: HiRes-2; open triangles:
      Hires-1; open diamonds: Fly's Eye. Fit of the Galactic
      (low-energy dot-dash line) and extra-galactic (high-energy
      dot-dash) to cosmic ray data. Total flux: solid line; dotted
      lines: upper $75^{\rm th}$ and lower $25^{\rm th}$ percentiles 
      for the prediction of
      the extra-galactic flux.}
      \label{fig:flux-total}
\end{figure}
\smallskip

\section{Results and discussion}  

Figure~\ref{fig:flux-total} shows the total
spectrum (Galactic + Extra-galactic) compared to the data, assuming
continuously emitting sources with density $n=10^{-5}$Mpc$^{-3}$ and
spectral index $\gamma=2.6$. The solid and dot-dashed lines for the
extra-galactic show the median spectrum obtained over $500$
realizations of the source locations. For each realization the locations of 
the first hundred closest sources (i.e. within $\simeq 140\,$Mpc) were 
drawn at random, using a uniform probability law per unit volume; for 
farther sources, the continuous source approximation is valid and it was used
numerically. The upper and lower dotted curves show the $75^{\rm th}$
and $25^{\rm th}$ percentiles around this prediction, 
meaning that only 25\% of spectra
are respectively higher / respectively lower than indicated by these curves.
This uncertainty is related to the location of the 
closest sources, see below.

  Considering the difficulty of comparing different datasets, the fit
shown in Fig.~\ref{fig:flux-total} appears  satisfying. One
should also note that this fit uses a minimum number of free
parameters ($\gamma$ and $l_{\rm scatt}$ at $10^{17}\,$eV), in order
to consider the most economical scenario. As discussed below, there
are various ways in which one could extend the present analysis,
although this comes at the price of handling a larger number of
(unknown) parameters.

  In Fig.~\ref{fig:flux-total}, a straight dashed line was drawn
across the region $1.5\cdot10^{18}\rightarrow 8\cdot10^{18}\,$eV in which the
propagation is neither rectilinear nor diffusive. These limits were
found by comparing the diffusive and rectilinear spectra with the no
magnetic field spectrum. In this energy range the diffusive path
length becomes of the same order as the rectilinear distance at some
point during the particle history.  The diffusion theorem~\cite{AB04}
suggests that the flux in this intermediate region should follow the
no magnetic field spectrum (in which case it would dip $\sim10\,$\%
below the dashed line around $3\cdot10^{18}\,$eV).  This theorem rests
on the observation that integrating Eq.~(\ref{eq:diff}) for a
continuous distribution of sources over an infinite volume gives the
rectilinear spectrum Eq.~(\ref{eq:rect}). However the actual volume is
bounded by the past light cone; this is why the diffusive spectrum shuts
off exponentially at energies $\gtrsim 10^{18}\,$eV. The rectilinear
part shuts off at energies $\lesssim 7\cdot10^{18}\,$eV as the
maximal lookback time that bounds the integral of Eq.~(\ref{eq:rect})
decreases sharply. Hence one might expect a small dip in the spectrum
around $2-3\cdot10^{18}\,$eV. Interestingly the data is not
inconsistent with such a dip at that location. Monte Carlo simulations
of particle propagation are best suited (and should be performed) to
probe the spectrum in this region.

  The scattering length was assumed to scale as $l_{\rm scatt}\propto
E^2$, and its value at $10^{18}\,$eV was set here to $17\,$Mpc. 
This scaling of the scattering length is typical for particles with 
Larmor radius larger
than the coherence scale of the field, in which case $l_{\rm scatt}
\simeq l_{\rm c}(r_{\rm L}/l_{\rm c})^2$. Since the Larmor radius
$r_{\rm L} \simeq 1\,{\rm Mpc}\,(E/10^{18}\,{\rm eV})(B/10^{-9}\,{\rm
G})^{-1}$, one may expect this approximation to be valid. In effect,
$1\,$Mpc is a strict upper bound to the coherence length of a
turbulent inter-galactic magnetic field~\cite{WB99,AB04}, and
available numerical simulations indicate much smaller coherence
lengths~\cite{DGST} in clusters of galaxies. A value $l_{\rm c} \sim
10\,$kpc could also be expected if the inter-galactic magnetic field
is produced by galactic outflows. The above condition for $l_{\rm
scatt}$ corresponds to $B\sqrt{l_{\rm c}}\sim 2.5\cdot
10^{-10}\,$G$\cdot$Mpc$^{1/2}$ for an all-pervading magnetic
field. Hence, for $l_{\rm c} \sim 20\,$kpc, and $B\sim 2\cdot10^{-9}\,$G (in
order to obtain the correct scattering length at $10^{18}\,$eV), one
finds $r_{\rm L}\gtrsim l_{\rm c}$ for $E\gtrsim 3\cdot10^{16}\,$eV.

  It is possible that the scaling of $l_{\rm scatt}$ with energy
changes in the range $10^{16}\rightarrow10^{17}\,$eV as $r_{\rm L}$
may become smaller than $l_{\rm c}$. There is no universal scaling for
$l_{\rm scatt}$ when $r_{\rm L}<l_{\rm c}$ as the exact relationship
then depends on the structure of the magnetic field; for instance, in
Kolmogorov turbulence, one finds $l_{\rm scatt} \propto r_{\rm L}$ for
$0.1 l_{\rm c} \lesssim r_{\rm L} \lesssim l_{\rm c}$ and $l_{\rm
scatt}\propto r_{\rm L}^{1/3}$ at lower energies~\cite{CLP01}. The
possible existence of regular components of extra-galactic magnetic
fields may also modify $l_{\rm scatt}$. A change in the scaling of
$l_{\rm scatt}$ with $E$, if it occurs at $E\gtrsim 10^{16.5}\,$eV,
would imply a different value for $B\sqrt{l_{\rm c}}$, with the
difference being a factor of order unity to a few. It is exciting to
note that, in the present framework, experiments such as
KASCADE-Grande~\cite{KASCADE} may allow to constrain the energy
dependence of the scattering length (hence the magnetic field
structure) by measuring accurately the energy spectrum and composition
between the first and second knees.

\begin{figure}[t]
      \centering
      \includegraphics[width=0.5\textwidth,clip=true]{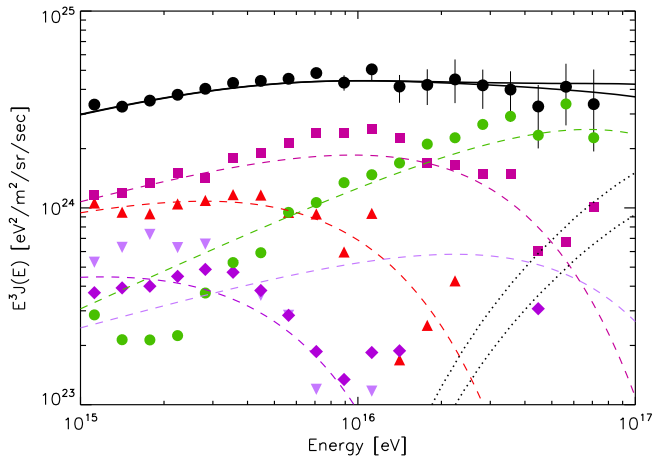}
      \caption[...]{KASCADE (SYBILL) data: solid lines: all-particle
      spectra (Galactic + extra-galactic); dotted lines:
      extra-galactic component; dashed lines: Galactic spectra, for p,
      He, C, Si and Fe, in order of increasing $x-$axis
      intercept. Filled circles: all-particle; diamonds: p;
      upward-pointing triangles: He; squares: C; downward pointing
      triangles: Si; circles: Fe. Error bars on reconstucted chemical
      composition have been omitted for clarity, but are significant,
      see ~\cite{KASCADE}.}  \label{fig:flux-knee}
\end{figure}

 The predictions (for both normalizations in
Fig.~\ref{fig:flux-total}) for the extra-galactic proton flux are shown
and compared to the chemical composition measurement of KASCADE in
Fig.~\ref{fig:flux-knee}. These composition measurements remain
uncertain, as can be seen by comparing the QGSJet and SYBILL
reconstructions in~\cite{KASCADE}; the proton and helium knee
positions seem robust however. The dotted lines represent the median 
proton signal from the extra-galactic component, whose detection seems
within the reach of KASCADE-Grande.  One may note that Galactic
spectra with exponential suppression beyond the knee agree with the
KASCADE data. Nonetheless, if the Galactic spectra are found to extend
as powerlaws beyond the knee, the scattering length of extra-galactic
protons should be smaller by a factor of order unity (and
$B\sqrt{l_{\rm c}}$ correspondingly higher).

  The result for $B\sqrt{l_{\rm c}}$ depends weakly on the source
density: since the diffusive (low energy) part of the spectrum shuts
off as $\exp\left[-r^2/4\lambda^2\right]$ with $r\sim n^{-1/3}$ the closest 
source distance, the cut-off energy depends on the ratio 
$n^{-1/3}/\lambda\propto n^{-1/3}(B\sqrt{l_{\rm c}})$, hence 
$B\sqrt{l_{\rm c}}$ scales with $n$
according to: $B\sqrt{l_{\rm c}}\sim 2-3\cdot10^{-10}(n/10^{-5}\,{\rm
Mpc}^{-3})^{1/3}\,$G$\cdot$Mpc$^{1/2}$.

Cosmic variance related to the distance $d$ to the closest sources 
is significant for the low energy ($E\lesssim 10^{17}\,$eV) and for
the high energy ($E\gtrsim 10^{20}\,$eV) parts of the
spectrum, as illustrated by the confidence intervals around the median
flux shown in Fig.~\ref{fig:flux-total}. In these two energy ranges, the
effective linear distance to the source is limited to $\lesssim
50\,$Mpc, which is comparable to the
expected distance to the closest source. Interestingly, the spectra
close to both low and high energy cut-offs are strongly correlated due to the above
effect. The distances to within which one should find 
$N=(1,2,3,4,5)$ sources (with $N$ the Poisson average) are
$r\simeq (29,36,41,46,50)\,$Mpc respectively. The diffusive spectrum 
sums up contributions that scale as
$\exp\left[-r_i^2/4\lambda^2\right]$, with $r_i$ the distance to the
$i^{\rm th}$ closest source. Therefore, close to the cut-off energy, 
where $r_1^2/4\lambda^2\gg 1$, the total spectrum is 
dominated on the average by the individual spectrum of the
closest or the two closest sources. At higher energies
 spectra of more remote sources contribute with a weight
$\sim\exp\left[-r_i^2/4\lambda^2\right]$. Since what matters most for
the comparison to the data is the cut-off energy, one finds that as a
first approximation, cosmic variance related to the position of the
closest source at distance $r_1$ induces an uncertainty of the
inferred magnetic field strength $\Delta B/B \sim \Delta d/d \sim {\cal
O}(1)$ since, as before, the cut-off energy depends on the ratio 
$r_1/\lambda$.

  It is possible that the ultra-high energy cosmic ray sources are
intermittent with an activity timescale $T_{\rm source}\ll H_0^{-1}$; the previous
discussion has assumed steady sources corresponding to $T_{\rm source}\sim
H_0^{-1}$. If $T_{\rm source}\ll H_0^{-1}$, the number density of
sources inferred from clustering at high energies ($E\gtrsim 4\cdot 10^{19}\,$eV) 
underestimates the actual density of {\it potential} sources by a
factor $T/H_0^{-1}$, with 
$T^2\sim T_{\rm source}^2 + \Delta\tau_B^2$, and where $\Delta\tau_B$
is the typical time spread at $E\sim4\cdot10^{19}\,$eV due to magnetic
delay. The average time delay reads: $\tau_B\sim 1.5\cdot 10^8\,{\rm yr}\,
(E/10^{19}\,{\rm eV})^{-2}(d/1\,{\rm Gpc})^2
(B\sqrt{l_{\rm c}}/3\cdot10^{-10}\,{\rm G}\cdot{\rm
Mpc}^{1/2})^2$~\cite{WM96}; the ratio $\Delta\tau_B/\tau_B\lesssim 1$
depends  on the structure of the random magnetic field, see~\cite{WM96,Lea97}.
Each source then contributes for a fraction $T/H_0^{-1}$ of a Hubble
time to the diffusive spectrum given in Eq.~(\ref{eq:diff}), but there
are $H_0^{-1}/T$ times more sources: the total flux remains
the same than evaluated previously, except that the cut-off energy
will correspond to that expected for a source density larger by 
$H_0^{-1}/T$. Hence, following the previous discussion,
the present scenario remains valid if the magnetic field strength
is higher by a factor $(H_0^{-1}/T)^{1/3}$. For instance, for active
galactic nuclei sources of ultra-high energy cosmic rays with $T_{\rm source}\sim
10^8\,$yr, a fit similar to that shown in Fig.~\ref{fig:flux-total}
can be obtained for a
magnetic field strength $B\sqrt{l_{\rm c}}\sim 10^{-9}\,{\rm
G}\cdot{\rm Mpc}^{1/2}$. For the particular case of 
transient {\em Galactic} sources, such as $\gamma-$ray bursts, the
situation is different, since the closest sources lie at distance
$r_i\approx0$. Therefore the
diffusive spectrum $J_{\rm diff}(E)\approx \sum_i (c/4\pi) 
\left[4\pi D(E)t_i\right]^{-3/2} q(E)$, since $\lambda^2\simeq
D(E)t_i$ for close by sources, 
with $t_i$ the lookback time to the $i^{\rm th}$ event, and 
$q(E)$ the injection spectrum per source. Diffusion in extra-magnetic 
fields thus does not produce a low energy cut-off in this case; the 
spectrum is rather subject to the fluctuations of the time
distribution of past Galactic events.

  At high energies, $E\gtrsim 10^{19}\,$eV, particles travel in
a  quasi-rectilinear fashion, i.e. the deflection angle suffered by
crossing a coherence cell of the magnetic field $\delta\theta\sim
l_{\rm c}/r_{\rm L}\sim 3\times10^{-3}(l_{\rm c}/30\,{\rm kpc})
(E/10^{19}\,{\rm eV})^{-1}(B/10^{-9}\,{\rm G})$ is much smaller than
unity. The total deflection angle summed over the trajectory remains
smaller than unity, and this justifies the use of Eq.~(\ref{eq:rect})~\cite{WM96}:
$\theta_{\rm rms}\sim 25^o (E/10^{19}\,{\rm eV})^{-1}(d/1\,{\rm
Gpc})^{1/2}(B\sqrt{l_{\rm c}}/3\cdot10^{-10}\,{\rm G}\cdot{\rm
Mpc}^{1/2})$. This also implies  that charged particle astronomy will
be possible at the highest energies. Recent studies have
attempted to obtain definite predictions for $\theta_{\rm rms}$ by
using MHD simulations of large-scale structure formation with magnetic
fields scaled to reproduce existing data in clusters of
galaxies~\cite{DGST,SME}.  Their results differ widely, thereby
illustrating the difficulty of constraining {\it ab initio} the
strength of extra-galactic magnetic fields. The present value for
$\theta_{\rm rms}$ is comparable to or slightly larger than that of
Ref.~\cite{DGST}, and substantially smaller than that of
Ref.~\cite{SME}. The magnitude of $\theta_{\rm rms}$ indicates that
extra-galactic magnetic fields could be probed through the angular
images of ultra-high energy cosmic ray point sources, and this will
constitute a strong test of the present scenario.

  The proposed scattering length cannot result from scattering on
magnetic fields associated with galaxies or groups and clusters of
galaxies, since the collision mean free path with either of these
objects is too large, being $\sim {\cal O}(1\,{\rm Gpc})$. The
inferred magnetic field might in principle be concentrated around the
source (on distance scale $L$) and negligible everywhere else. Since
the spectrum would cut off below an energy such that $2\lambda\approx
2\left[cH_0^{-1}l_{\rm scatt}\right]^{1/2} \lesssim L$, this requires
$B\gtrsim 1\,\mu{\rm G}\, (l_{\rm c}/10\,{\rm kpc})^{-1/2}
(L/100\,{\rm kpc})^{-1}$ (for a cut-off at $10^{17}\,$eV). 
This possibility cannot be excluded but it
gives a non-trivial constraint on the source environment. Searches for
counterparts at the highest energies would help test this possibility:
for instance, magnetic fields such as above are found in clusters of
galaxies but there is no report of clusters in the arrival directions
of the highest energy events. If this magnetic field is intrinsic to
the source, or if the cut-off at $\lesssim10^{18}\,$eV is due to
injection physics in the source~\cite{BGG2}, then, under the present
assumptions, the present work still gives a stringent upper bound on
all-pervading magnetic fields. To remain conservative, one may require
that the cut-off should not occur above $\sim 10^{18}\,$eV, in which
case one finds $B\sqrt{l_{\rm c}}\lesssim
10^{-9}\,$G$\cdot$Mpc$^{1/2}$. This limit is still an order of
magnitude below existing Faraday bounds.

  The magnetic field in question thus appears inter-galactic in
nature, in which case it is likely to be inhomogeneously distributed
on small scales. Further studies are then required to relate the
average $B\sqrt{l_{\rm c}}$ with the actual structure and distribution
of these magnetic fields. One needs to account for the possible
existence of a regular magnetic field component aligned with filaments
and walls, which would inhibit perpendicular transport, and consider
the respective filling fractions and amplitudes of the turbulent and
regular components. It would be certainly worthwhile to extend the
simulations of particle propagation made in realistic magnetic
fields~\cite{DGST,SME} to the energies of interest.

  Finally there are various ways in which the present study could be
extended. One should notably consider the possible energy dependences
of the scattering length (including the above effects of inhomogeneous
and regular magnetic fields), the role of intermittent sources, the
possible cosmological evolution of the magnetic field and of the
source density, and, as mentioned above, the possibly inhomogeneous 
structure of the magnetic field on a scale comparable to the closest
ultra-high energy cosmic ray sources.

\begin{acknowledgments}
A referee is acknowledged for a constructive report; it is a pleasure
to thank ``Le S\'eminaire'' for hospitality where part of this work
was performed.
\end{acknowledgments}

\appendix*
\section{Diffusion over cosmological scales}
The diffusion of particles in an expanding background space-time can
be seen as a standard diffusion process on a fixed background
in conformal coordinates $(\eta,{\mathbf r})$, with $\eta$ the
conformal time defined by $a(\eta) {\rm d}\eta={\rm d}t$ and ${\mathbf
r}$ the comoving coordinates in a Friedman-Lema\^\i
tre-Robertson-Walker metric; $a$ denotes the scale factor and $t$
cosmic time. One can indeed approximate the diffusing process as a random
walk against scattering centers of constant comoving coordinates. 

  Particles also experience dilution due to expansion,
expansion energy losses and energy losses due to pair and pion
production on diffuse backgrounds. 
At redshift $z=0$, these photo-interaction losses are  negligible
with respect to expansion losses for energies
$E\lesssim 2\cdot10^{18}\,$eV, but become increasingly
more important at higher
redshift due to the increased cosmic microwave background temperature and
density~\cite{BGG1}. Nonetheless the main energy loss in the course of
the history of a particle with present energy  $E_0\lesssim
10^{18}\,$eV is due to expansion. One reason is that the  majority
of  the sources that
contribute to the diffuse flux at energy $E_0$ are located at moderate
redshifts as a result of the nonlinear time-redshift relation: 
redshift $z=2$, for instance, corresponds to a lookback time of $76$\%
of the age of the Universe~\cite{cosmology}. More importantly, 
pion and pair production losses at high redshift become catastrophic, 
so that the time interval
during which the losses are dominated by photo-interactions
 is much smaller than a Hubble
time. Finally, the contribution to the diffuse flux at energy $E_0$ of 
particles injected with energy $E_{\rm g}$ scales, in
a first approximation, as $q(E_{\rm g}) {\rm d}E_{\rm g}/{\rm d}E$,
with $q(E_{\rm g})$ the injection spectrum and ${\rm d}E_{\rm g}/{\rm
d}E$ acounts for the dilation of the energy interval. 
The function $E_{\rm g}(\eta,E)$ defines the energy of
the particle at time $\eta$, assuming it has energy $E$ at time $t_0$.  
This function and its derivative ${\rm d}E_{\rm g}/ {\rm d}E$ can be 
reconstructed by integrating the energy losses~\cite{BGG1}. Assuming
${\rm d}E_{\rm g}/{\rm d}E \approx E_{\rm g}/E$, which is exact for 
expansion losses, one sees that the contribution of particles injected
at remote lookback times (hence with high $E_{\rm g}$) is negligible
with respect to that of particles injected recently with $E_{\rm
g}\approx E$ since $q(E_{\rm g})\propto E_{\rm g}^{-\gamma}$ and
$\gamma\sim 2.6$ here. The numerical difference between a diffuse flux
computed using only expansion losses and that computed with all energy
losses included is indeed less than $5$\% at $E\ll 10^{18}\,$eV, and increases to
$20$\% at $E\sim 10^{18}\,$eV. Consequently, 
it is assumed in this discussion that particles
with present $E_0\lesssim 10^{18}\,$eV have been subject to expansion losses
only throughout their history. 

Energy losses due to expansion are
expressed as: ${\rm d}E/{\rm d}\eta = -{\cal H}E$, with ${\cal
H}\equiv (1/a) {\rm d}a/{\rm d}\eta$ the expansion rate in conformal time.
 The phase space density of particles $N(\eta,E,{\mathbf r})$ at
coordinates ${\mathbf r}$, time $\eta$ and energy $E$, which is
related to the distribution function by 
$N(\eta,E,{\mathbf r})=(4\pi p^2/c)f(\eta,{\mathbf p},{\mathbf r})$, 
with $pc=E$, is solution to the diffusion equation:
\begin{equation}
{\partial\over\partial\eta}\left(a^3N\right) - \nabla D\nabla
\left(a^3N\right) -{\cal H}{\partial\over\partial E}\left(E
a^3N\right)\,=\,a^3\tilde Q\left(\eta,E,{\mathbf r}\right),
\label{eq:diff-eq}
\end{equation}
where $\tilde Q$ gives the (physical) number density of particles
injected per unit energy and conformal time intervals.
 The $a^3$ prefactor of $N$ takes
into account the effect of dilution of particle density through
expansion. The fact that expansion losses are separable in terms of 
the two variables $E$ and $\eta$ allows  to find an exact solution
to this diffusion equation, using standard Green functions methods;
see~\cite{S59} for the same problem with time independent losses in a
non-expanding background. Explicitly, through the change of variables
$(\eta,E)\rightarrow (u,v)$, with $u=\log(aE)$ and $v=\log(a/E)$, one
can derive the Green function (for the equation for $N$) as:
\begin{eqnarray}
G\left(\eta_0,E_0,{\mathbf r}_0;\eta_e,E_e,{\mathbf r}_e \right)&
\,=\, & \left({a_e\over a_0}\right)^2
{\exp\left[-{\vert{\mathbf r_0-\mathbf r_e}\vert^2\over
4\lambda^2}\right]\over\left(4\pi\lambda^2\right)^{3/2}}\nonumber\\
& & \times\delta\left(E_e-{a_0E_0\over a_e}\right),
\label{eq:Green}
\end{eqnarray}
with the shorthand notations: $a_0\equiv a(\eta_0)$ and $a_e\equiv
a(\eta_e)$. The path length $\lambda$ is defined by:
\begin{equation}
\lambda^2\,=\, \int_{\eta_e}^{\eta_0} {\rm d}\eta\, D\left[{a_eE_e\over
a(\eta)}\right],\label{eq:lambda}
\end{equation}
where $D(E)$ is the diffusion coefficient; if this latter depends
explicitly on time, for instance if the magnetic field strength
evolves with redshift, the solution remains valid.


\begin{thebibliography}{99}
\bibitem{cosmology} It is assumed that $\Omega_{\rm m}=0.3$,
$\Omega_\Lambda=0.7$, $H_0=70\,$km/s/Mpc.
\end{thebibliography}
 \end{document}